# Piercing an interface with a brush : collaborative stiffening

F. Chiodi, B. Roman & J. Bico

*Physique et Mécanique des Milieux Hétérogènes,*
*UMR 7636 CNRS ESPCI-Paris6-Paris7*
*ESPCI, 10 rue Vauquelin 75005 Paris*

**Abstract.** - The hairs of a painting brush withdrawn from a wetting liquid self-assemble into clumps whose sizes rely on a balance between liquid surface tension and hairs bending rigidity. Here we study the situation of an immersed carpet in an evaporating liquid bath : the free extremities of the hairs are forced to pierce the liquid interface. The compressive capillary force on the tip of flexible hairs leads to buckling and collapse. However we find that the spontaneous association of hairs into stronger bundles may allow them to resist capillary buckling. We explore in detail the different structures obtained and compare them with similar patterns observed in micro-structured surfaces such as carbon nanotubes "forests".

**Introduction.** – Everyday's life experience teaches us that wet hairs assemble into bundles. This phenomenon is however amplified at the scale of Micro-Electro-Mechanical-Systems (MEMS) since surface forces tend to dominate over bulk forces when the scale is reduced. Indeed, if $L$ is the typical size of a structure, surface forces are proportional to $L$, while elastic or gravity forces scale as $L^2$ and $L^3$, respectively. Controlling 'stiction' is then a challenging issue in micro-engineering technologies since it often leads to the fatal collapse of microstructures [6–8]. Nevertheless, the self-assembly of micro-structures through capillary forces can also be viewed as a useful tool to build complex shapes [1–5]. Beyond engineering applications, surface forces may also have a strong effect on living structures. For instance, filamentous fungi living in aqueous environment have difficulty in growing their hypha through the water interface into the air. Indeed some species have to produce surfactant molecules that reduce capillary forces in order to develop the aerial structures necessary for dissemination [24].

In the case of slender structures, the interaction between elasticity and interfacial forces can be defined by a typical *elastocapillary* length scale, $L_{EC} = \sqrt{B/\gamma} \sim \sqrt{Eh^3/\gamma}$, where $E$ is the Young modulus of the material, $h$ and $B$ are the thickness and the bending stiffness per unit width of the structure, respectively, and $\gamma$ the liquid surface tension or the solid adhesion energy [25–29]. The validity of this macroscopic length scale has recently been confirmed at the scale of graphene sheets through atomistic simulations [30].

In this paper we study the case of a carpet-like structure immersed in a drying liq-uid bath. This situation is important for practical situations in microtechnologies since microstructures are often dried out of a solvent and brought to pierce the liquid interface during the evaporation process. Recent experiments with wet carbon nanotubes [9–13], ZnO [14] or Si [15–17] nanorods 'carpets' and polymeric micro-pilars arrays [18–22] exhibit a surprising large variety of bundle structures ranging from 'tepee' shapes to cellular patterns. Surprising helicoidal structures have also recently been observed with soft PDMS carpets [23]. However no attempt has been made to classify the different regimes.

Although macroscopic studies have shown that an isolated structure buckles upon capillary forces if its length is larger than a critical length of the order of $L_{EC}$ [29], little is known about the collective piercing (or collapsing) of an assembly of bristles. May the structure assemble to resist the capillary forces and pierce the interface? The aim of this paper is to present a configuration diagram that predicts the final equilibrium states as a function of the length of the bristles, their rigidity and their lattice spacing. We will first extend the results on the formation of bundles, study the piercing of isolated bundles, and finally deduce a configuration diagram. Although our study is limited to macroscopic regular 1D brushes we believe that our results are relevant to the scale of nanotubes 'carpets'.

**Experimental setup.** – 1D model brushes are build by clamping lamellae of length $L$ (centimetric) cut from bi-oriented polypropylene sheets (Innovia Films, $E \simeq 2\,\mathrm{GPa}$) of thickness $h$ (of 15, 30, 50 or 90 $\mu$m) on a base with a regular spacing $d$ (ranging from millimeters to centimeters). The elestocapillary length $L_{EC}$ is measured for each thickness with the 'racket' technique described by Py et al. [1]. The brushes are first immersed into a bath of commercial dish-washing solution that totally wets the lamellae ($\gamma = 26.5\,\mathrm{mM/m}$). In order to mimic evaporation, the liquid is progressively drained from the reservoir, which brings the free ends of the lamellae in contact with the liquid/air interface (fig. 1). Successive images from a typical experiment are displayed in figure 2. As the tips reach the interface, lamellae tend to merge spontaneously into bundles. Depending on the experimental parameters, bundles pierce the interface without much damage or instead buckle and eventually collapse (fig. 3).

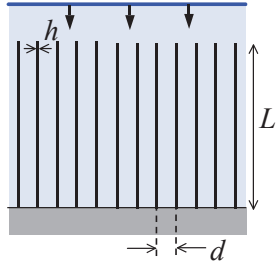

Fig. 1: Sketch of the experiment. Lamellae of length $L$ and thickness $h$ are clamped on an immersed base with a regular inter-spacing $d$. As the liquid is progressively removed from the reservoir, the free tips of the lamellae are forced to pierce the liquid/air interface.

**Forming bundles.** – We first present the sticking of wet lamellae by capillary forces, once out of the liquid bath. When a macroscopic brush with a regular lattice is withdrawn (tips down) from a liquid bath, pairs of intermediate bundles successively stick together, leading to large hierarchical bundles [26, 28] (in this situation, the lamellae do not have to pierce the interface; we will consider later if these bundles are stable in the inverted case). A balance between adhesion and elastic bending energy gives the distance from the root $L_{stick}$ at which two hairs with an initial spacing $d$ join [1]. In the limit of small deformations

---
[1]This argument is similar to the classical argument derived by Obreimoff in 1930 to estimate the splitting strength of mica [31]

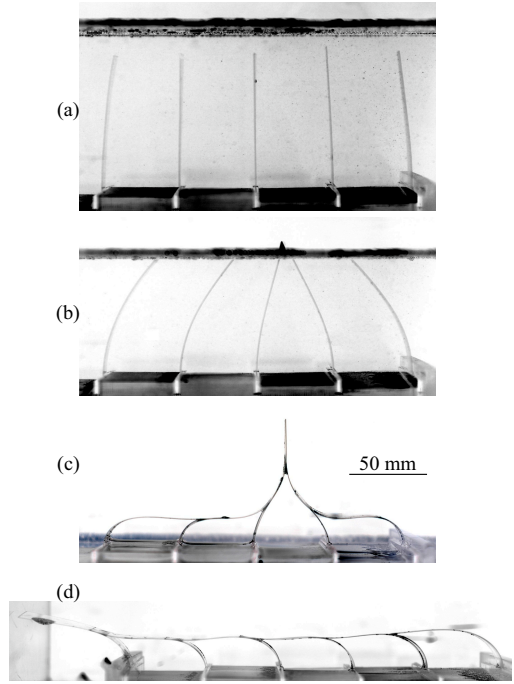

Fig. 2: Typical experiment: (a) immersed brush, (the upper dark line corresponds to the liquid surface); (b) as the liquid is progessively removed, the interface reaches the tips of the lamellae, isolated lamellae buckle and eventually collapse but may also bundle together and pierce the liquid surface; (c) final bundle ($L = 90$ mm, $d = 50$ mm, $L_{EC} = 33.6$ mm); (d) Another experiment in the same configuration leads to the collapse of all the lamellae.

($d/L_{stick} \ll 1$), the solution is analytical [26, 27]:

$$L_{stick} = \left(\frac{9}{2}\right)^{1/4} (dL_{EC})^{1/2}. \tag{1}$$

This relation can be extrapolated to intermediate bundles of size $N/2$ merging into larger bundles of size $N$, by multiplying the bending stiffness by factor $N/2$ (we assume that the liquid prevents friction between lamellae), and using an effective distance between these intermediates bundles $Nd/2$. This leads to an effective elasto-capillary length of $(N/2)^{1/2} L_{EC}$, so that the joining length $L_{stick}$ of a pair of intermediate bundles merging into a bundle of size $N$ is in this case given by [26]:

$$L_{stick}(N) = \frac{\sqrt{3}}{2} N^{3/4} (dL_{EC})^{1/2}. \tag{2}$$

We see that the formation of large bundle require long lamellae ($L_{stick}$ increases with $N$). Conversely, for a given brush with lamellae of length $L$, the maximum size of a bundle $N_{max}$ is easily derived from this last equation by taking $L_{stick} = L$:

$$N_{max} = 2\left(\frac{2}{9}\right)^{1/3} \left(\frac{L^4}{d^2 L_{EC}^2}\right)^{1/3}. \tag{3}$$

However smaller bundles generally are also present: when the sum of the sizes of neighboring bundles exceeds $N_{max}$, they cannot stick together because they would lead to a bundle exceeding $N_{max}$. When a brush is withdrawn bundles are randomly formed. The statistical

F. Chiodi, B. Roman & J. Bico

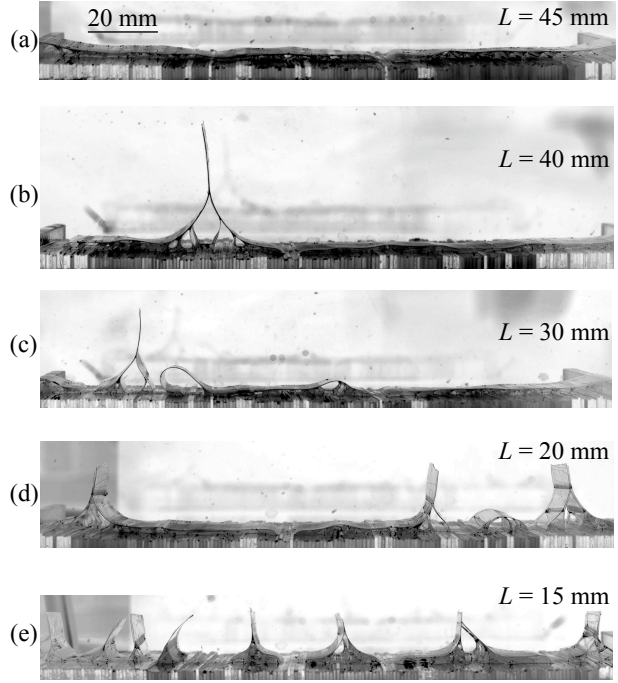

Fig. 3: Final state for decreasing lamellae lengths ($d = 5\,\text{mm}$, $L_{EC} = 5.5\,\text{mm}$). All lamellae collapse for the longest samples, while an increasing number of bundles pierce the liquid surface without any damage as the length is reduced.

distribution of the size of the bundles is found to follow a self-similar size law [32]. This peculiar distribution has a maximum size $N_{max}$ and also a minimal size, which is of the order of $N_{min} \simeq 0.3 N_{max}$.

The condition of small deformations ($d/L_{stick} \ll 1$) assumed for deriving the previous relations is however not always verified in practice (*e.g.* experiment displayed in fig. 1). The equilibrium shape of the lamellae can nevertheless be described in the general situation by solving numerically Euler's *elastica* relation [33]:

$$B\frac{d^2\theta}{ds^2}\mathbf{e_z} + \mathbf{t} \times \mathbf{R} = \mathbf{0}, \qquad (4)$$

where $\theta$ is the angle made by the tangent to the lamella $\mathbf{t}$ with the vertical at the curvilinear coordinate $s$, $\mathbf{e_z}$ the vector perpendicular to the plane and $\mathbf{R}$ the constant vectorial tension of the beam (in the present case, $\mathbf{R}$ only has an horizontal component). Most boundary conditions required to solve the equation are trivial: $\theta = 0$ at the contact point and at the clamped end, the horizontal displacement is equal to $d/2$ between the two points. The last boundary condition is less obvious and can be derived from a balance between elastic and surface energy [1]: the curvature at the contact point is $\sqrt{2}/L_{EC}$. The non-dimensional form of eq. 4 was solved numerically for increasing values of the non-dimensional distance $d/L_{EC}$. The corresponding ratio $L_{stick}/(dL_{EC})^{1/2}$ is displayed in fig. 4 and is found to be fairly well fitted by a linear correlation: $L_{stick}/(dL_{EC})^{1/2} \simeq (9/2)^{1/4}(1 + 0.043 d/L_{EC})$. We can finally extrapolate this relation to the case of a pair of intermediate bundles of size $N/2$ (which implies an effective elasto-capillary length of $(N/2)^{1/2} L_{EC}$, and an effective distance $Nd/2$) merging into a bundle of size $N$:

$$L_{stick}(N) \simeq \frac{\sqrt{3}}{2} N^{3/4} (dL_{EC})^{1/2} \left(1 + 0.030\sqrt{N}\frac{d}{L_{EC}}\right). \qquad (5)$$

Solving this relation for $L_{stick} = L$ then gives the maximum size $N_{max}$ corresponding to the non-dimensional spacing $d/L_{EC}$. Once we have characterized the size of the bundles that can spontaneously form on a given brush, we now wonder if these elastic structures may pierce the liquid interface, or buckle and collapse.

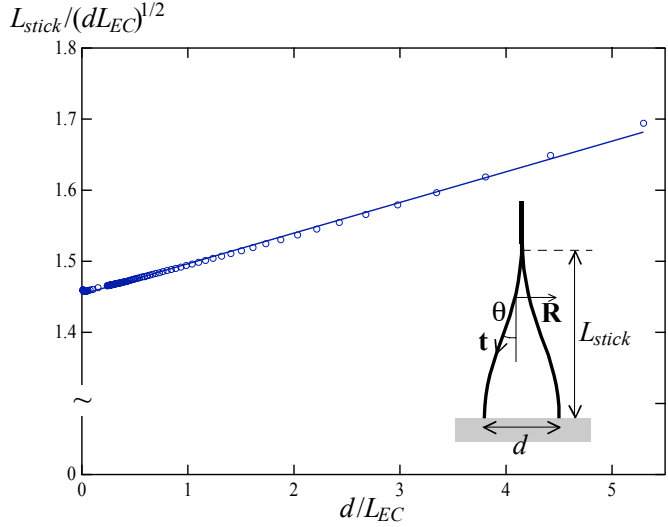

Fig. 4: Sticking length of a pair of hairs accounting for the finite value of the spacing $d$. Symbols: numerical solutions of the *elastica* equation. Line: linear correction of the zero-order relation (eq. 5), $L_{stick} = (9/2)^{1/4}(dL_{EC})^{1/2}(1 + 0.043d/L_{EC})$

**Piercing an interface with an isolated bundle.** – We first consider the case of an isolated lamella that pierces the liquid surface (insert in fig. 5). If the liquid wets the material, the vertical capillary force pushing the lamella downward at the liquid interface is given by: $F_{cap} = 2w\gamma$, where $w$ is the width of the lamella [2]. Classical Euler buckling criterion predicts a critical length for the slender lamella above which it buckles:

$$L_{crit} = \frac{\pi}{2}\sqrt{\frac{Bw}{2\gamma w}} = \frac{\pi}{2\sqrt{2}} L_{EC}, \quad \text{with} \quad B = \frac{Eh^3}{12(1-\nu^2)}, \tag{6}$$

where $E$ and $\nu$ are the material Young modulus and Poisson ratio, respectively, and $h$ the thickness of the lamellae. Although the actual postbuckling behavior is more complex [29], we will suppose, for the sake of simplicity, that lamellae with lengths exceeding $L_{EC}$ eventually collapse towards the base.

In the case of a brush, the buckled lamellae generally hit their neighbors and merge into larger bundles. May these more rigid structures resist capillary loading and pierce the interface? Since the liquid can lubricate the relative displacement between lamellae, we would expect a bundle involving $N$ lamellae to be $N$ times stiffer than a single lamella, leading to an increase of $L_{crit}$ by a factor $\sqrt{N}$. The minimum piercing size $N_0(L)$, above which a bundle of a given length $L$ is strong enough to resist piercing is given by:

$$N_0(L) = \frac{8}{\pi^2}\left(\frac{L}{L_{EC}}\right)^2. \tag{7}$$

---

[2]We suppose here that the contact angle is zero, otherwise $\gamma$ should simply be replaced by $\gamma\cos\theta$, where $\theta$ is the contact angle of the liquid on the surface

In order to compare this theoretical prediction with the results obtained with our brushes separated with a spacing $d$, we measured the minimum size $N_{crit}(d, L)$ above which a bundle of artificially fixed size resits. $N_{crit}(d, L)$ is found significantly lower than the predicted value $N_0(L)$ as the spacing $d$ is increased (fig. 5). We qualitatively interpret this result by the larger width of the base of the bundle, which increases the effective stiffness of the structure. Experimental data obtained with lamellae of different lengths and thicknesses collapse in a single master curve when $N_{crit}/N_0$ is plotted as a function of $d/L$. A fair fit of this master curve is (full line in fig. 5):

$$N_{crit}(d, L) = \frac{N_0(L)}{1 + 16d/L}. \tag{8}$$

As the liquid bath is drained, lamellae tend to merge into bundles. Bundles containing more than $N_{crit}$ lamellae pierce the interface while the other ones collapse.

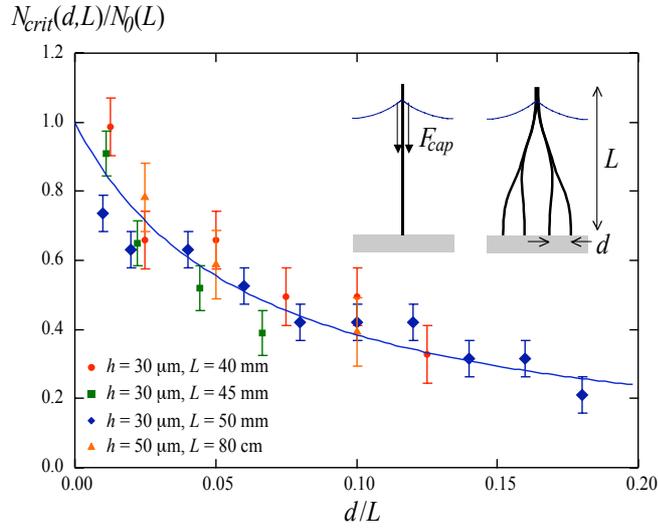

Fig. 5: Main figure: minimum size of a bundle of a given length required to pierce the liquid interface as a function of the distance between lamellae. Continuous line: empirical fit $N_{crit}(d, L)/N_0(L) = 1/(1 + 16d/L)$. Inset: compressive capillary force acting on a single lamella or on a bundle.

**Configuration diagram.** – The fate of a given brush can be predicted by studying the piercing conditions of the bundles that it spontaneously develops. This is done by comparing the critical size $N_{crit}$ with the maximum and minimum bundle sizes $N_{max}$ and $N_{min}$. Indeed, if for a given length $L$, the size $N_{crit}$ (eq. 8) exceeds $N_{max}$ (eq. 3), the whole brush is expected to collapse (case 1 in fig. 6). If $N_{crit}$ lies between $N_{max}$ and $N_{min}$, lamellae merge into bundles as the liquid is removed; the largest bundles should pierce the interface while the smaller ones should collapse (case 2). If $N_{crit}$ becomes lower than $N_{min}$ even the smallest bundles are expected to pierce the surface of the liquid (case 3). A last situation arises for small values of $N_{max}$. When $N_{max}$ is lower than 2, lamellae do not form bundles and can either remain straight if $L < L_{crit}$ (case 4) or otherwise collapse (case 5, which joins case 1). The different cases are summarized in table 1 and sketched in fig. 6. These predicted regimes are in good agreement with experiments displayed fig. 2 and fig. 3.

Although our model brushes are one-dimensional, we expect our results to be qualitatively valid for two-dimensional 'carpets'. As a main quantitative difference, the scaling with $N^{3/4}$ in eq. 3 becomes $N^{3/8}$ [28]. The direct comparison with experiments carried with nanorods is not precise since $L_{EC}$ was nor measured, but provides some qualitative indication. For instance, cellular patterns similar to case 2 are observed in the experiments described by Chakrapani *et al.* [11]. In these experiments, the radius of the multi-wall nanotubes $b$ is assumed to be of the order of 15 nm and the material Young modulus $E \sim 1$ TPa, which gives $L_{EC} \sim 0.2\,\mu$m (in the case of rods $L_{EC} = \sqrt{\pi E b^3/4\gamma}$ [28]). The length of the tubes is much larger than $L_{EC}$ ($L \sim 100\,\mu$m), which favors the collapsed cases 1 and 2, while the lattice spacing $d \sim 0.05\,\mu$m may not be small enough to prevent collapse. Conversely, 'tepee' structures reminding case 3 are formed in the experiments described by Lau *et al.* [10], with $b \sim 25$ nm, giving $L_{EC} \sim 0.4\,\mu$m of the same order of magnitude as the length $L \sim 4\,\mu$m and the lattice spacing ($d \sim 0.3\,\mu$m). Experiments with Si rods exhibit the same structure [15] with $b \sim 20$ nm, $E \sim 130$ GPa, leading to $L_{EC} \sim 0.1\,\mu$m for a length $L \sim 1\,\mu$m and a spacing $d \sim 0.13\,\mu$m. If these nanorods were isolated they would not have pierce the interface but instead buckle and eventually collapse. This study suggests that collaborative piercing is possible for arbitrary flexible structures if they can merge into large enough bundles.

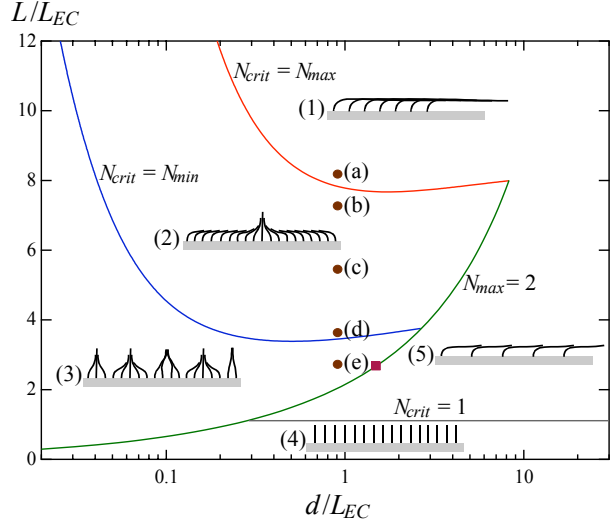

Fig. 6: Configuration diagram and comparison with experiments. Case 1: bundles of any possible size collapse. Case 2: the largest bundles resist an pierce the surface, while smaller one collapse. Case 3: Bundles of all accessible sizes resist. Case 4: lamellae do not form bundles but are stiff enough to pierce the surface. Case 5: lamellae do not form bundles and collapse (this case joins Case 1). Experimental parameters corresponding to the experiments illustrated in fig. 2 (■) and fig. 3 (●).

Note finally that other theoretical approaches have been proposed in the literature. The formation of bundles is then interpreted in terms of lateral interactions [11,13,15,22], which is basically equivalent to the aggregation described without the piercing problem. In addition, the finite thickness of the hairs becomes important when the distance between the hairs is small ($h \sim d$) and has been also considered [17]. However, to the best of our knowledge, the combination of the size distribution of the bundles with possible buckling is original.

Table 1: Different cases described in the configuration diagram.

| case 1 | $N_{max} < N_{crit}$ | the whole brush collapses |
|---|---|---|
| case 2 | $N_{min} < N_{crit} < N_{max}$ | biggest bundles pierce, while smaller collapse |
| case 3 | $N_{crit} < N_{min}$ | bundles of any size pierce |
| case 4 | $N_{max} < 2$ ; $L < L_{crit}$ | lamellae do not form bundles and remain straight |
| case 5 | $N_{max} < 2$ ; $L_{crit} < L$ | lamellae do not form bundles and collapse |

**Conclusion.** – The fate of a brush immersed in a drying liquid bath is determined by two competitive interfacial phenomena: compressive capillary forces may induce the buckling and eventually the collapse of the bristles, while lateral attractive capillary forces lead to collaborative stiffening through the formation of bundles. Different final states have been observed with model experiments on macroscopic brushes depending on the physical parameters of the brush. We showed that these physical parameters can be condensed into two non-dimensional parameters: $L/L_{EC}$ and $d/L_{EC}$, where $L$ is the length of the hairs, $d$ their spacing and $L_{EC}$ an *elastocapillary* length comparing bending stiffness to surface forces. Dense brushes of rigid hairs tend to resist capillary forces while floppy hairs in scarce brushes collapse. We found that arbitrary flexible hairs (that would collapse as individuals), may develop a collaborative sitffening by sticking to their close-enough neighbors and manage to pierce the interface. Although our study is limited to one-dimensional brushes, we believe that our results are qualitatively valid for two-dimensional situations and may help designing 'hairy' microstructures.

∗ ∗ ∗

This work was partially founded by the Société des Amis de l'ESPCI. We thank Dominic Vella and Guillaume Batot for fruitful discutions.